\documentclass[final]{svjour2}
\usepackage{graphicx}
\usepackage{rotating}
\usepackage{amssymb}
\usepackage{mathptmx}
\usepackage[numbers]{natbib}
\usepackage{color}

\usepackage{url}

\newcommand{\mytilde}{\raise.17ex\hbox{$\scriptstyle\mathtt{\sim}$}} 
\newcommand{\Planck}{{\it Planck}}   

\newcommand  \gtsim  {\lower.5ex\hbox{$\; \buildrel > \over \sim \;$}}
\newcommand  \ltsim  {\lower.5ex\hbox{$\; \buildrel < \over \sim \;$}}

\def\3he{$^3{\rm He}$}

\def\lsim{\mathrel{\lower2.5pt\vbox{\lineskip=0pt\baselineskip=0pt
           \hbox{$<$}\hbox{$\sim$}}}}

\def\gsim{\mathrel{\lower2.5pt\vbox{\lineskip=0pt\baselineskip=0pt
           \hbox{$>$}\hbox{$\sim$}}}}

\makeatletter
\journalname{Journal of Low Temperature Physics}

\bibpunct{}{}{,}{s}{}{,}

\begin{document}

\newcommand{\hdblarrow}{H\makebox[0.9ex][l]{$\downdownarrows$}-}
\title{Measuring Reionization, Neutrino Mass, and Cosmic Inflation with BFORE}

\author{Sean Bryan$^1$ \and Peter Ade$^2$ \and J. Richard Bond$^3$ \and Francois Boulanger$^4$ \and Mark Devlin$^5$ \and Simon Doyle$^2$ \and Jeffrey Filippini$^6$ \and Laura Fissel$^7$ \and Christopher Groppi$^1$ \and Gilbert Holder$^6$ \and Johannes Hubmayr$^8$ \and Philip Mauskopf$^1$ \and Jeffrey McMahon$^9$ \and Johanna Nagy$^{10,11}$ \and C. Barth Netterfield$^{11}$ \and Michael Niemack$^{12}$ \and Giles Novak$^{13}$ \and Enzo Pascale$^{14}$ \and Giampaolo Pisano$^2$ \and John Ruhl$^{10}$ \and Douglas Scott$^{15}$ \and Juan Soler$^{16}$ \and Carole Tucker$^2$ \and Joaquin Vieira$^6$}

\institute{$^1$School of Earth and Space Exploration, Arizona State University, Tempe, AZ USA\\
\email{sean.a.bryan@asu.edu}\\
$^2$School of Physics and Astronomy, Cardiff University, Cardiff, UK\\
$^3$Canadian Institute for Theoretical Astrophysics, University of Toronto, Toronto, ON, Canada\\
$^4$Institut d'Astrophysique Spatiale, Orsay, France\\
$^5$Department of Physics and Astronomy, University of Pennsylvania, Philadelphia, PA, USA\\
$^6$Department of Physics, University of Illinois at Urbana-Champaign, Urbana, IL, USA\\
$^7$National Radio Astronomy Observatory, Charlottesville, NC, USA\\
$^8$National Institute of Standards and Technology, Boulder, CO, USA\\
$^9$Department of Physics, University of Michigan, Ann Arbor, MI, USA\\
$^{10}$Department of Physics, Case Western Reserve University, Cleveland, OH, USA\\
$^{11}$Department of Astronomy and Astrophysics, University of Toronto, Toronto, ON, Canada\\
$^{12}$Department of Physics, Cornell University, Ithaca, NY, USA\\
$^{13}$Department of Physics and Astronomy, Northwestern University, Evanston, IL, USA\\
$^{14}$Department of Physics, Sapienza Universit\`{a} di Roma, Rome, Italy\\
$^{15}$Department of Physics and Astronomy, University of British Columbia, Vancouver, Canada\\
$^{16}$Max Planck Institute for Astronomy, Heidelberg, Germany}
\maketitle

\begin{abstract}

BFORE is a NASA high-altitude ultra-long-duration balloon mission proposed to measure the cosmic microwave background (CMB) across half the sky during a 28-day mid-latitude flight launched from Wanaka, New Zealand. With the unique access to large angular scales and high frequencies provided by the balloon platform, BFORE will significantly improve measurements of the optical depth to reionization $\tau$, breaking parameter degeneracies needed for a measurement of neutrino mass with the CMB. The large angular scale data will enable BFORE to hunt for the large-scale gravitational wave $B$-mode signal, as well as the degree-scale signal, each at the $r\sim0.01$ level. The balloon platform allows BFORE to map Galactic dust foregrounds at frequencies where they dominate, in order to robustly separate them from CMB signals measured by BFORE, in addition to complementing data from ground-based telescopes.  The combination of frequencies will also lead to velocity measurements for thousands of galaxy clusters, as well as probing how star-forming galaxies populate dark matter halos. The mission will be the first near-space use of TES multichroic detectors (150/217 GHz and 280/353 GHz bands) using highly-multiplexed mSQUID microwave readout, raising the technical readiness level of both technologies.


\end{abstract}

\section{Introduction}

The temperature and polarization anisotropies of the Cosmic Microwave Background (CMB) encode a wealth of information about reionization, cosmic inflation, and the large-scale structure of the universe. Precision large-scale structure measurements with the CMB are moving towards a detection of the sum of the neutrino masses $m_\nu$, a major outstanding question in particle physics. These and other science goals\citep{CMBS4-ScienceBook} are the context for the recently funded Simons Observatory, the proposed CMB-S4 project, as well as the current generation of ground-based \citep{henderson/etal:2015,suzuki16,essinger-hileman/etal:2014,benson/etal:2014,grayson16} and balloon-borne \citep{fraisse13,gandilo16,ebex17a} CMB instruments.


\begin{figure}
\begin{center}
\includegraphics[width=0.65\linewidth]{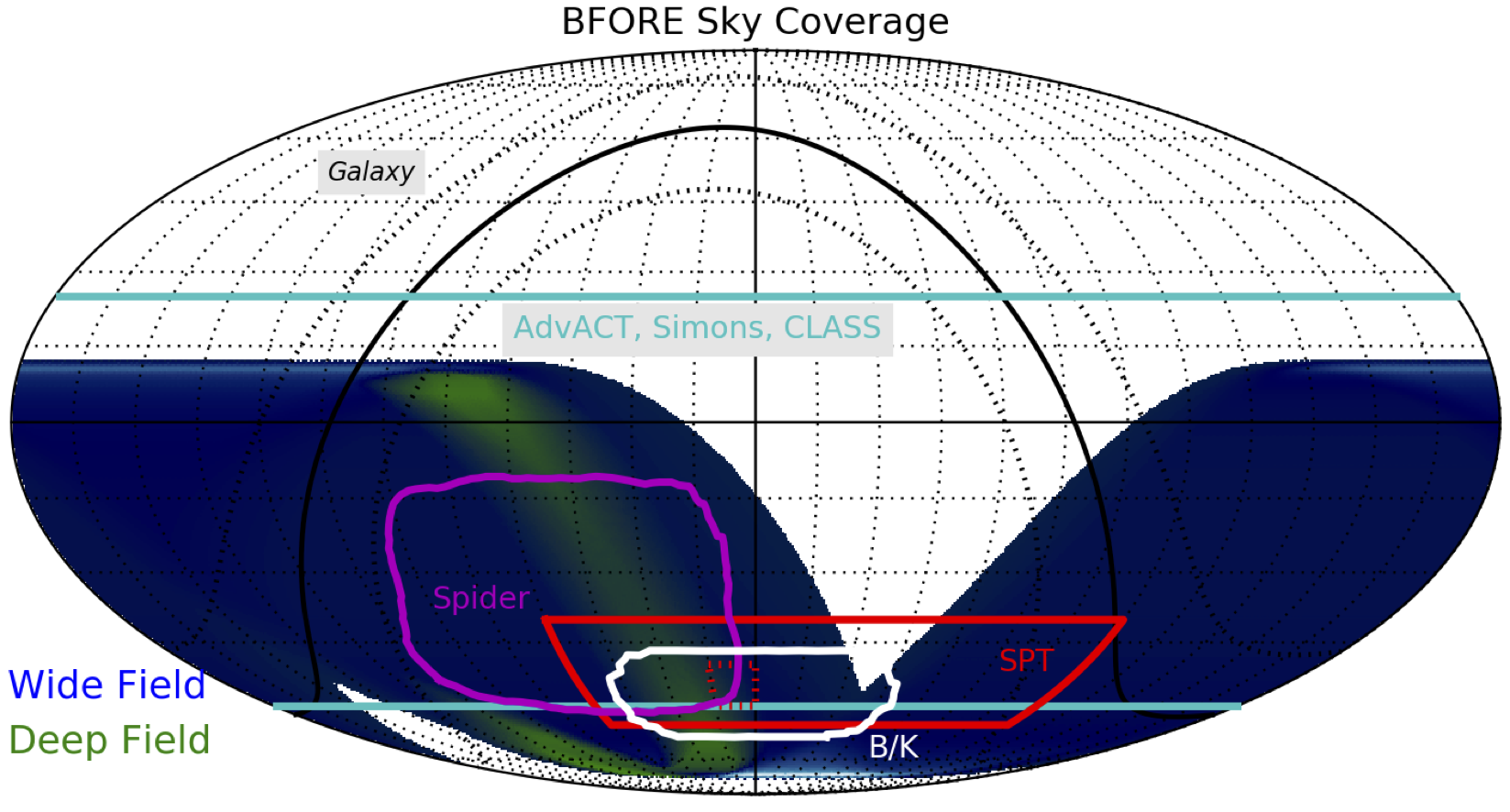}
\caption{\label{fig:sky} Simulated sky coverage for a 28-day BFORE ULDB flight launched from New Zealand on April 15th, 2021. The wide field is shown in blue in the top panel and covers $50\%$ of the sky, overlapping with the Simons Observatory and the current generation of ground-based CMB experiments, \citep{henderson/etal:2015,suzuki16,essinger-hileman/etal:2014,benson/etal:2014,grayson16} as well as the Spider balloon instrument. \citep{fraisse13,shariff15} The 2500\,deg$^2$ deep field (green) is away from the Galactic plane (black solid and dashed lines).}
\end{center}
\end{figure}

BFORE is a proposed CMB polarization experiment designed to be flown on a NASA ultra-long duration balloon (ULDB) launched from Wanaka, New Zealand. A key advantage of ULDB, compared with conventional NASA Antarctic flights, is a dramatic increase in available sky coverage. During the nighttime portions of the flight, BFORE will spin to map $50\%$ of the sky (20,000\,deg$^2$, shown in Figure~\ref{fig:sky}) in four bands (150, 217, 280, and 353~GHz). This sky coverage, as well as the stability of the TES detectors and balloon platform, will allow the instrument to reconstruct the large angular scale modes needed to make a significantly improved measurement of the optical depth to reionization $\tau$, breaking degeneracies needed for a detection of neutrino mass. The data will also set an improved upper limit on large angular scale gravitational wave $B$ modes from inflation. The similar angular resolution and overlapping sky coverage with current and upcoming ground-based CMB experiments means that the higher frequency bands of BFORE will play an important role in dust foreground cleaning. During the daytime portions of the flight, the instrument will map a 2{,}500\,deg$^2$ deep field to set an independent upper limit on neutrino mass and degree-scale gravitational wave $B$ modes, and also to study lensing, galactic dust, galaxy cluster properties and star formation.

\section{Science Goals}

\begin{figure}
\begin{center}
\includegraphics[width=0.67\linewidth]{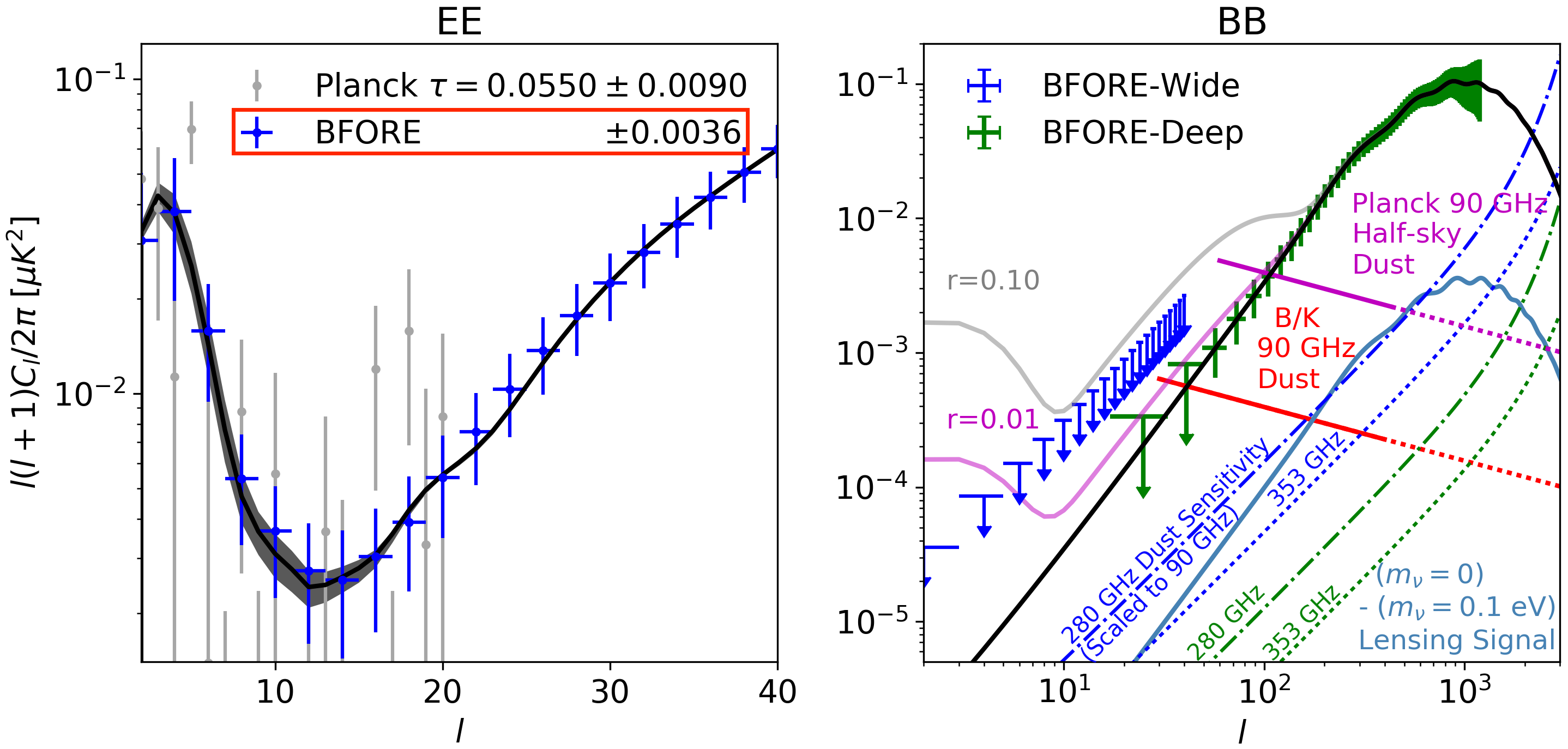}
\includegraphics[width=0.319\linewidth]{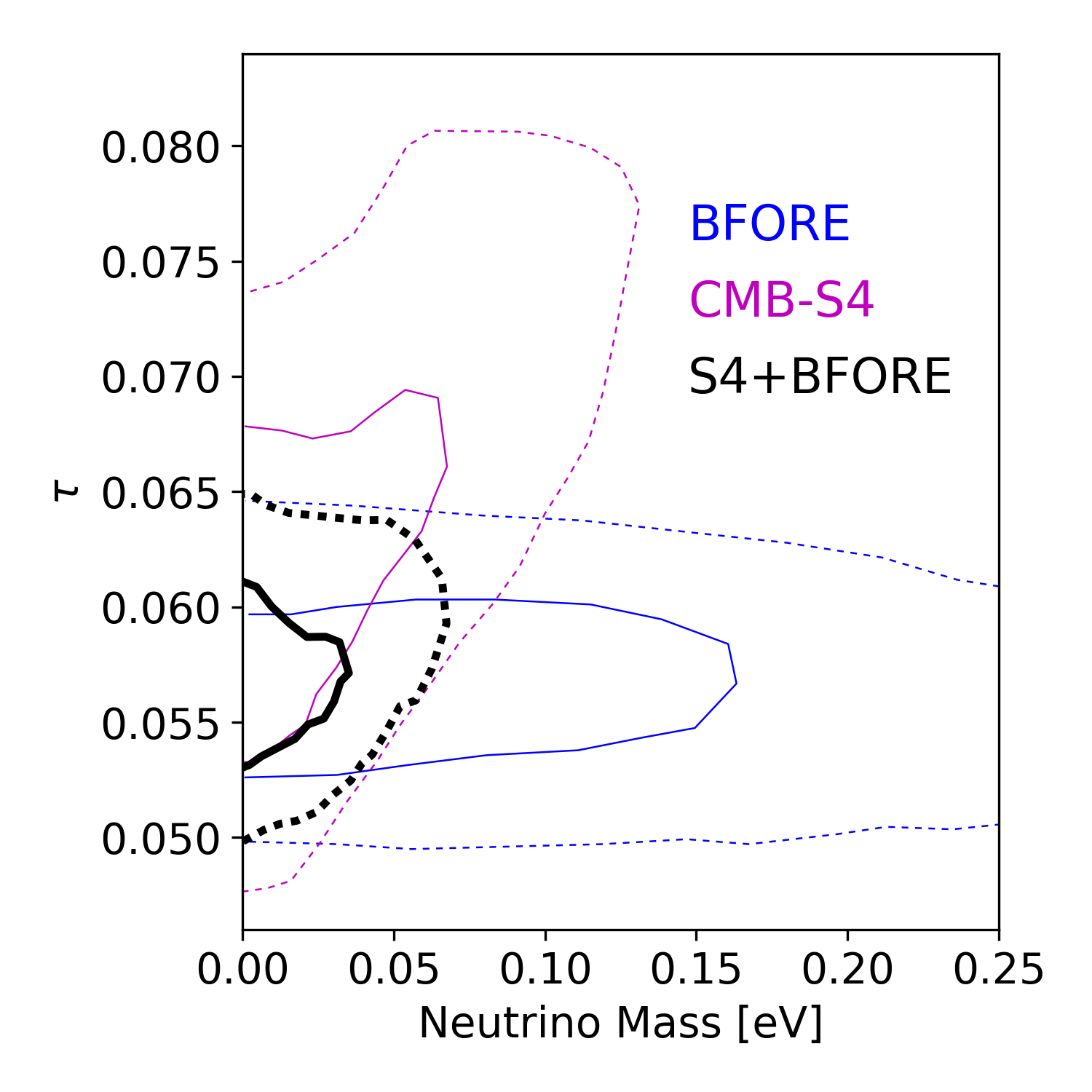}
\caption{\textbf{ Left:} BFORE $E$-mode data alone will improve the error on the optical depth to reionization $\tau$ by nearly a factor of 3 over the \textit{Planck} 2016 $\tau$ measurements. \citep{planck2016-XLVI} \textbf{Center:} The BFORE half-sky $B$-mode data (blue points) will constrain the large angular scale gravitational wave signal, and the 2,500\,deg$^2$ deep field data (green points) will constrain the degree-scale signal, each at the $r\lesssim0.01$ level. BFORE data alone will measure neutrino mass to 0.166\,eV (1$\sigma$), and will yield foreground data over half the sky at 280\,GHz (dashed blue and green lines) and 353\,GHz (dotted blue and green lines) of unprecedented sensitivity. \textbf{Right:} Ground-based CMB measurements that lack sensitivity to large angular scales due to atmospheric fluctuations suffer from a degeneracy between $\tau$ and neutrino mass. The BFORE $\tau$ measurement will break this degeneracy. Forecasting\citep{CMBS4-ScienceBook} shows that Simons Observatory and/or CMB-S4, $\tau$ constraints better than available from \textit{Planck} but \textit{at the level BFORE will provide}, plus DESI-like BAO measurements, are all necessary to detect the minimal 0.058~eV neutrino mass at more than 3$\sigma$. \label{fig:forecast}}
\end{center}
\end{figure}

\subsection{Neutrinos and Reionization}

The CMB is sensitive to neutrino mass because when neutrinos are relativistic they free stream and damp the clustering of matter. Their mass therefore imparts a scale at which the gravitational lensing power spectrum is suppressed. This lensing signal is reconstructed from the CMB temperature and polarization maps using $B$-modes and higher-order statistics. However, as illustrated in Figure \ref{fig:forecast}, in ground-based CMB measurements limited by atmospheric fluctuations to angular scales smaller than a few degrees, there is a degeneracy between neutrino mass and $\tau$. This arises because the apparent amplitude of lensed $B$-modes depends on the neutrino mass, but it also depends on the amplitude of primordial perturbations $A_{\rm s}$. However, small-angular-scale CMB measurements observe the damped perturbations, meaning they measure the parameter combination $A_{\rm s} e^{-2 \tau}$. This causes a degeneracy\citep{hou/etal:2013, manzotti/etal:2015} between $\tau$ and the lensed $B$-mode amplitude, and therefore also a degeneracy with neutrino mass.

In 2016, the \Planck\ team published constraints on reionization with improved treatment of large angular scale systematics and noise, yielding $\tau = 0.0550 \pm 0.0090$ from the {\tt SimBaL} simulation-based estimator\citep{planck2016-XLVI}, and $\tau = 0.0580 \pm 0.0120$ from the {\tt Lollipop} cross-spectrum estimator. \citep{planck2016-XLVII} These uncertainties on $\tau$ leave enough of a degeneracy to preclude\citep{CMBS4-ScienceBook} a 3$\sigma$ detection of the minimal neutrino mass, even if a future ground-based experiment like CMB-S4 achieves an 1~$\mu$K-arcmin map noise and even if CMB-S4 combines with DESI-like measurements of large-scale structure. This means that the current instrument-noise-limited precision of the \Planck\ $\tau$ constraints will limit the science reach of the Simons Observatory, CMB-S4, and potentially current generation experiments.

As illustrated in Figure \ref{fig:forecast}, BFORE will break this $\tau$--$m_\nu$ degeneracy by precisely measuring the CMB $E$-modes on large angular scales, yielding an independent measure of $\tau$ with nearly 3 times better error than \textit{Planck}. The projected BFORE error on $\tau$ of $\pm 0.0036$ is powerful enough to combine\citep{CMBS4-ScienceBook} with DESI and Simons Observatory (or CMB-S4) to yield a 3$\sigma$ or better detection of the minimal $0.058$~eV neutrino mass. The BFORE $\tau$ measurement will be within a factor of 2 of the all-sky CMB cosmic-variance limit of $\pm 0.0020$. In general, the angular scales relevant for $\tau$ ($\ell<20$, i.e. $\gtsim 9^\circ$) are difficult to access from the ground, especially at higher frequencies where atmospheric emission increases. Moreover, these angular scales are difficult to reach from the limited sky area available to Antarctic balloon-based measurements, showing that a ULDB mission like BFORE is uniquely powerful for this measurement. 

\begin{table}[bt]
 \vspace{-.35in}
\caption{BFORE Instrument Parameters and Calculated Map Depths} 
\label{tab:tel}  
\begin{tabular}{| l l l | }\hline \hline 
\textbf{Telescope:} &Temperature&250\,K (Primary), 4.2\,K (Secondary)\\ 
& Primary diameter&1.35\,m, emissivity $\le 0.005$\\ \hline
\textbf{Detectors:}
& Central frequencies  & {\bf 150~~~~217~~~~280~~~~353~GHz}\\ 
\textit{mSQUID} & Number of TESs & 2400~~2400~~2880~~2880 \\ 
	\textit{readout}  & Detector NEP & 6~~~~~~~~8~~~~~~~~10~~~~~~12 ~~$\mathrm{aW}/{\sqrt{{\rm Hz}}}$ \\ 
 & Background power~~ & 1.0~~~~~1.5~~~~~2.0~~~~~2.5 ~pW\\ 
\textit{100 mK fridge}& Background NEP~~ & 17~~~~~~23~~~~~~31 ~~~~~37 ~~$\mathrm{aW}/{\sqrt{{\rm Hz}}}$\\ \hline
\textbf{Half-sky map:}  & $\Delta Q/U$ (RJ) & 13~~~~~~11~~~~~~18~~~~~~12~~~$\mu$K-arcmin\\
\textit{28 nights} & $\Delta Q/U$ (CMB) & 22~~~~~~34~~~~~~101~~~~158 $\mu$K-arcmin\\
\textit{w/ 90\% obs. eff.}& $\Delta Q/U$ (Dust scaled to 150\,GHz) & 22~~~~~~10~~~~~~13~~~~~~9~~~~~$\mu$K-arcmin \\ \hline
\textbf{Deep map:} & $\Delta Q/U$ (RJ) & 4.1~~~~~3.6~~~~~5.7~~~~~3.8~~$\mu$K-arcmin\\
\textit{28 days}& $\Delta Q/U$ (CMB) & 7.1~~~~~11~~~~~~32~~~~~~54 ~~$\mu$K-arcmin\\
\textit{w/ 90\% obs. eff.}& $\Delta Q/U$ (Dust scaled to 150\,GHz) & 7.1~~~~~3.2~~~~~4.2~~~~~3.2~~$\mu$K-arcmin \\ \hline
\textbf{Beam:} & FWHM & 6.1~~~~~4.2~~~~~3.3~~~~~2.6~~arc-minutes \\ 
\hline 
 \end{tabular}
\end{table} 

\subsection{Inflation}

Cosmic inflation is a paradigm that naturally explains the observed gaussianity and scale-invariance of cosmological perturbations. A key additional prediction of inflation is a background of gravitational waves that would imprint a $B$-mode signal onto the CMB at roughly degree scales at recombination, and also at large angular scales at reionization. A major science goal of the current and next generation of CMB experiments is to definitively detect these inflationary $B$-modes. However, their amplitude (quantified by $r$) is constrained to be small; for example, GUT-scale inflation \citep{baumann09} suggests $r \sim 0.01$, which is the target for the current generation of ground-based CMB experiments.

As illustrated in Figure~\ref{fig:forecast}, BFORE will use its unique access to large angular scales from the ULDB platform to measure or constrain the large angular scale $B$-mode signal with $r\sim$0.01 sensitivity. It will also measure the degree scale signal to a similar level. Being above the atmosphere means that BFORE will measure galactic dust foreground emission in the 280- and 353-GHz bands to improve foreground removal from its own maps and complementary ground-based data. In the event of a detection of a potential inflationary gravitational wave signal by BFORE or another experiment, the large sky coverage of BFORE will be important in evaluating whether or not the signal is isotropic and scale-invariant, and the 280- and 353-GHz maps will be important for verifying that the signal is not contaminated by dust foregrounds.

\section{Instrument Approach}

The sky coverage needed to achieve the science goals of BFORE leads naturally to a mid-latitude flight. The current NASA ULDB program has a stricter weight budget than the Antarctic program, which the instrument has been designed to meet. The weight budget in turn limits available power from batteries and solar panels, and drives the design of the cryogenic receiver to an efficient liquid helium system. The instrument will use TES detectors with a highly multiplexed microwave SQUID readout. The gondola and pointing systems are based on high heritage systems flown recently on Spider.\citep{gandilo14,shariff15}

\begin{figure}
\begin{center}
\includegraphics[width=1.0\linewidth]{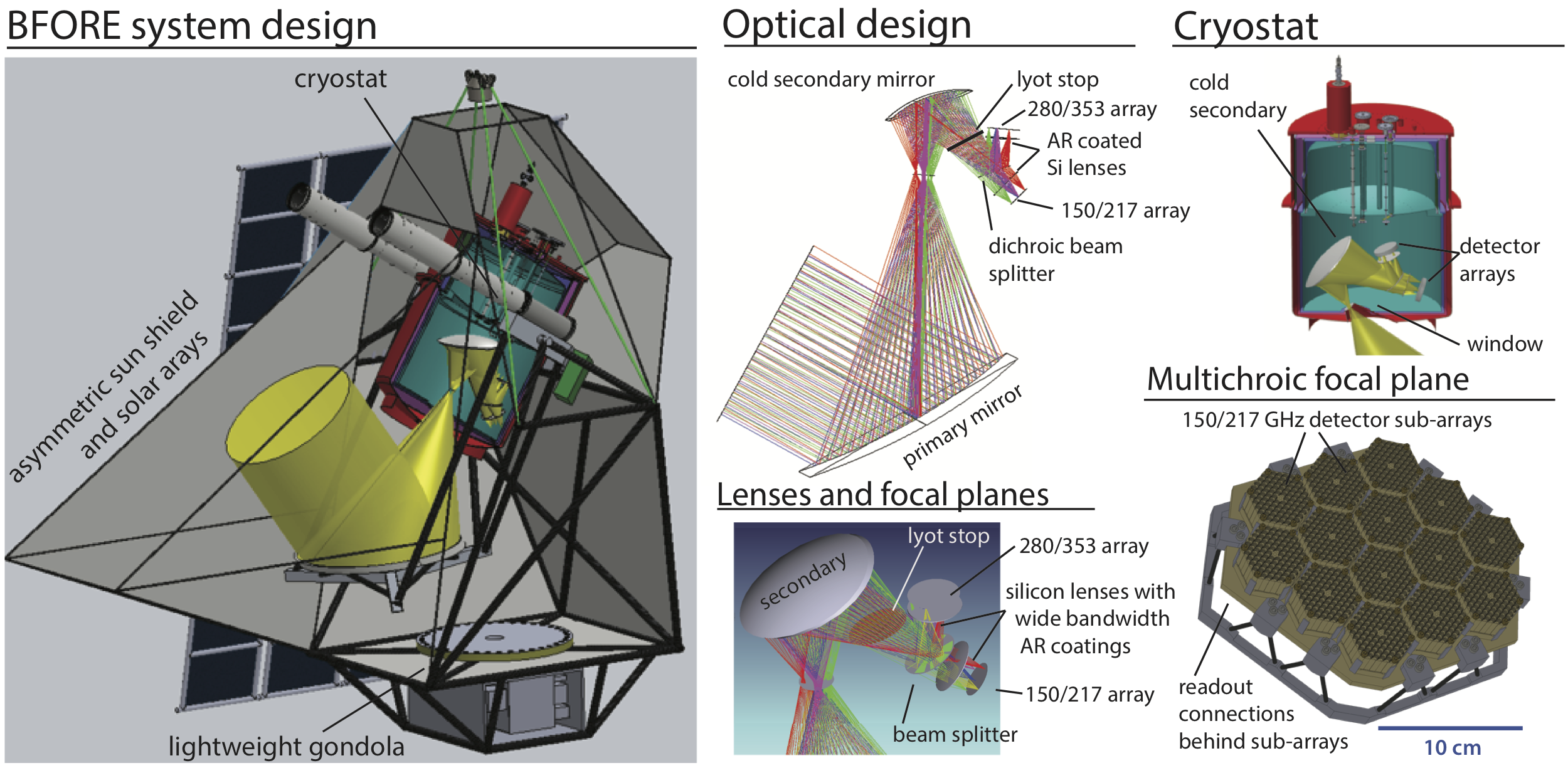}
\caption{ {\bf Left:}  BFORE gondola with solar panels, star cameras, and partially-shown sun shields. {\bf Middle:}  The BFORE optical design uses anti-reflection coated silicon lenses to illuminate a 1.35 m effective diameter on the primary mirror.  
{\bf Top Right:} Cutaway view of the cryostat showing the 4-K secondary mirror and 100-mK focal plane array.  The cryostat holds 500 liters of liquid helium.  {\bf Bottom~Right:}  The BFORE 150/217-GHz focal plane consists of 10 sub-arrays, each with 120 feed horns, for a total of 4800 TESs. The 280/353-GHz array has a similar sub-array architecture, but with 5760 TESs.}
\end{center}
\label{fig:optics}
\end{figure}

The optical system uses a single ambient temperature mirror to increase the angular resolution over what is feasible with all-cryogenic optics, while minimizing the loading contribution. Following the cryogenic secondary mirror and the Lyot stop, there is a dichroic to split the 150/217 and 280/353 bands, and a single cold lens forming an image onto each focal plane. The cryostat window has a diameter smaller than $20$\,cm, so we are able to use a thin polypropylene window. This reduces loading on the detectors and cryogenic system, and eliminates the need for a complex mechanical retractable window cover.

\begin{figure}
\begin{center}
\includegraphics[height=0.345\linewidth]{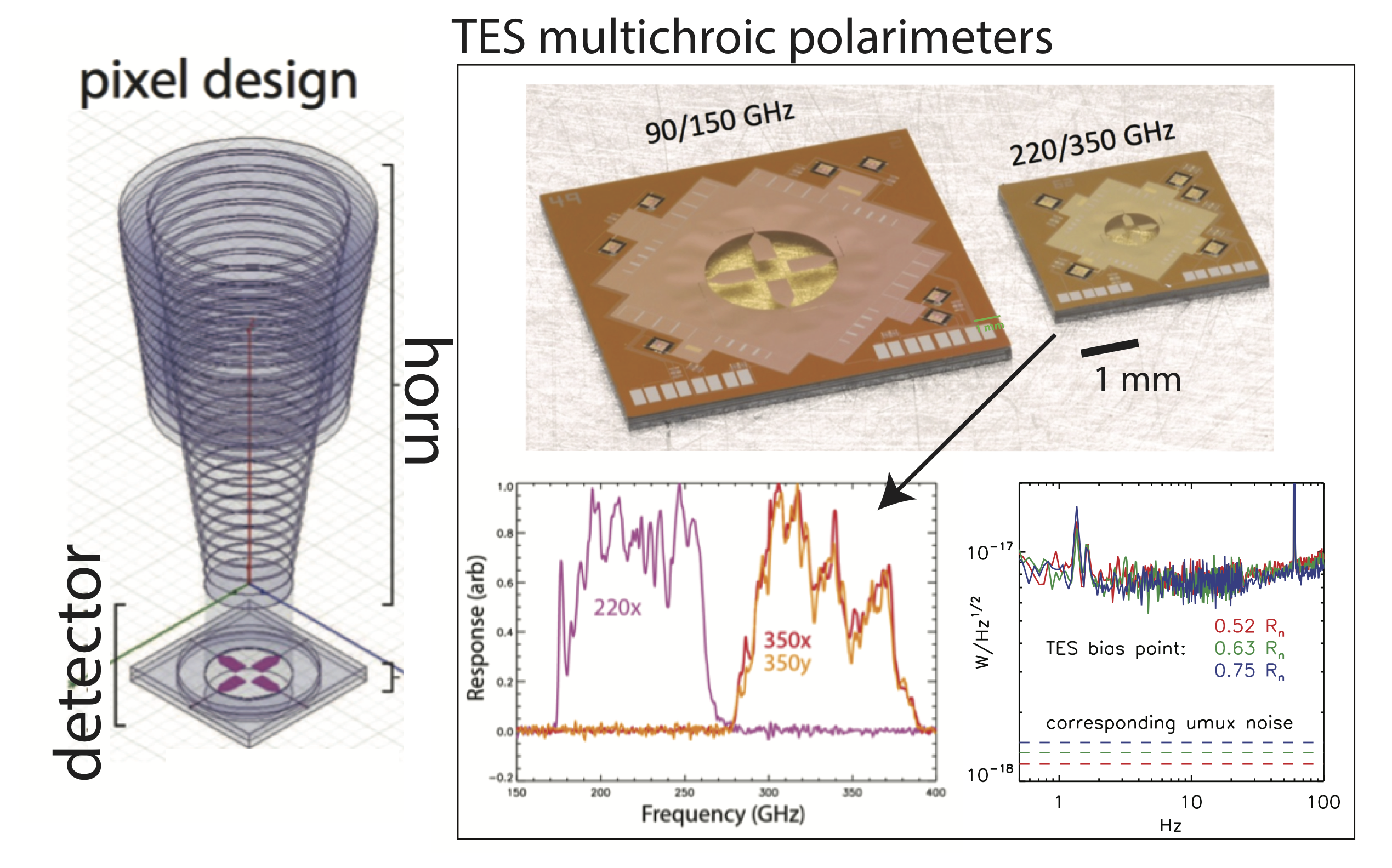}
\includegraphics[height=0.345\linewidth]{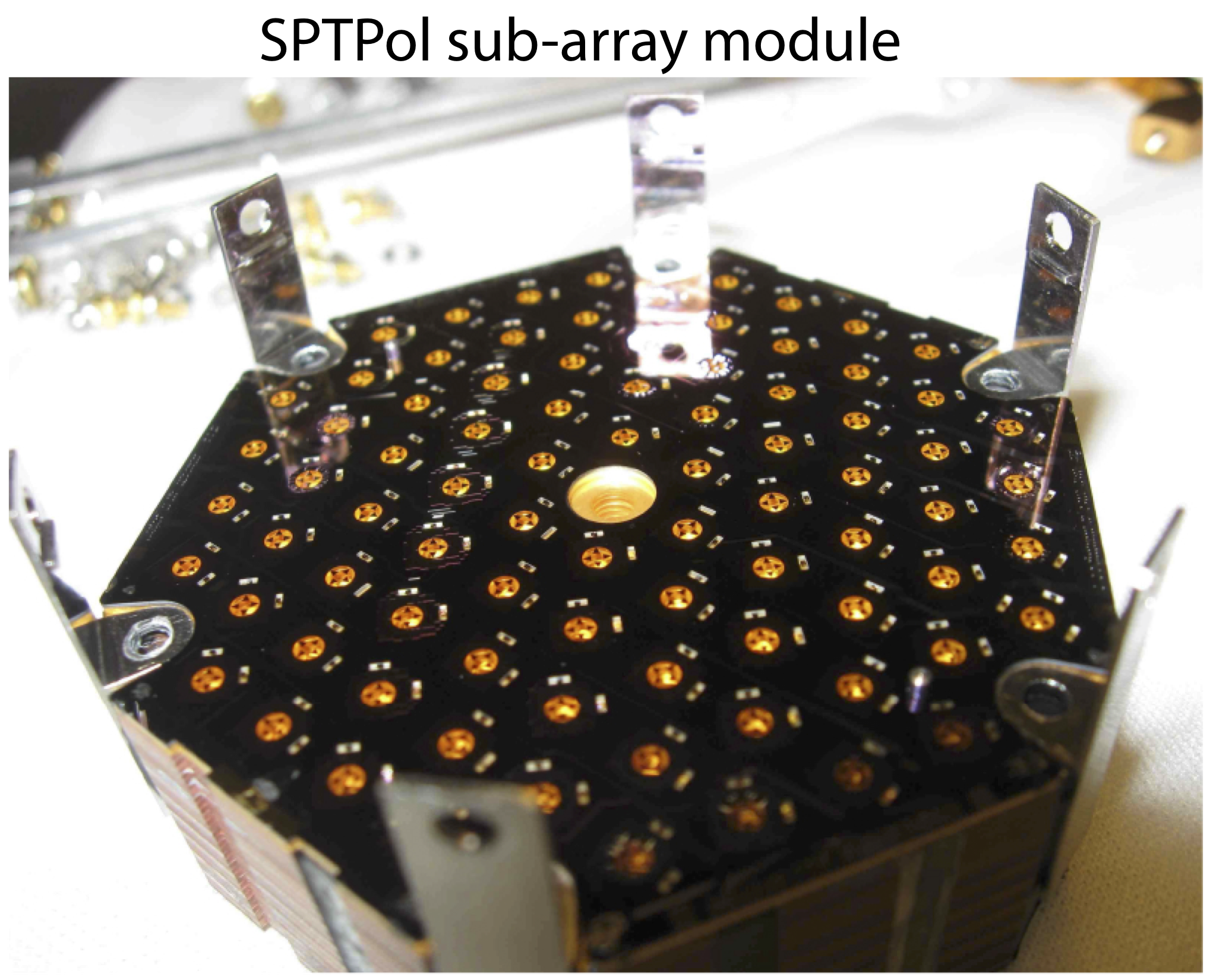}
\caption{{\bf Left/Middle:} Horn-coupled TES multichroic polarimeters at 90/150\,GHz and 220/350\,GHz, similar to those currently used in ACTPol and SPTPol. Measured 220/350-GHz passbands are also shown.  The noise measurement shows that the bolometers deliver better NEP than required for BFORE. \citep{niemack/etal:2012}  The current noise of the microwave mux has been measured to be 17 pA/$\sqrt{\mbox{Hz}}$, much lower than the TES noise. {\bf Right:} The full arrays depicted in Figure~\ref{fig:optics} consist of sub-arrays similar to this 84-feedhorn SPTPol module.\citep{henning/etal:2012} The photo shows the back side of the module with the detector wafer exposed before the backshort was installed on top of the wafer.}
\end{center}
\label{fig:detectors}
\end{figure}

The high-heritage NIST-fabricated multichroic detector arrays in BFORE will combine the proven performance of TES detectors with new low-dissipation mSQUID microwave readout. The arrays will be cooled to 0.1\,K to take full advantage of the low-background balloon environment. As illustrated in Figure~\ref{fig:detectors} and Table~\ref{tab:tel}, the detector arrays have a total of 10{,}560 bolometers across two arrays of multichroic horn-coupled polarization-sensitive pixels, operating in the 150/217 and 280/353-GHz bands. Similar dichroic pixels have been used in two seasons of observations on the ACTPol experiment with a 90/150 GHz array \citep{datta/etal:2014,datta/etal:2015} and one season (so far) with a 150/220-GHz array on Advanced ACTPol.\citep{henderson/etal:2015}

Polarization modulation is often used in CMB experiments to reduce the impact of detector $1/f$ noise and reduce the impact of some beam systematics. This comes at the expense and complexity of additional moving parts and payload weight. Also, if a modulator is used in a large diameter optical system such as BFORE, it is not practical to make the modulator the first optical element, which reduces its ability to mitigate systematics. The BFORE detectors and readout have demonstrated a low enough $1/f$ knee to reconstruct modes on the sky over an entire gondola rotation, meaning that as with Spider\cite{bryan16}, rapid modulation is not needed for BFORE. On degree and larger angular scales (for $\tau$ and primordial $B$-mode science), beam systematics should be negligible due to the small primary beam. For dust foreground mapping and other high $\ell$ science, the mid-latitude flight provides more than 45$^\circ$ of sky rotation, enough to completely switch $Q$ to $U$ without a modulator. Modern optimal mapmaker software \citep{bicep_III,louis16} uses this rotation between the sky and instrument to separate beam systematics from true sky signal. Since there are proven strategies to address detector stability and beam systematics without a modulator, to reduce payload weight and system complexity, BFORE will not include a modulation subsystem. The focal planes have interleaved $Q$ and $U$ pixels for further consistency checks on beam and polarimetric systematics, a technique used successfully by SPTPol. \citep{austermann_spt_2012}

\section{Conclusions}

Studying reionization to move towards a detection of neutrino mass, large angular scale $B$-modes from gravitational waves, and high frequency CMB dust foregrounds, are all important science goals that naturally point towards a mid-latitude ULDB balloon experiment such as BFORE. The payload architecture has been designed with the ULDB weight and power constraints in mind, with reductions in system complexity based on previous experience with balloon- and ground-based CMB measurements. Because the mission would be the first sub-orbital use of the multichroic TES pixels and microwave SQUID multiplexing, the flight would raise the NASA technical readiness level (TRL) of these technologies, paving the way for their use in future satellite programs.


\bibliographystyle{unsrt}
\bibliography{refsPLUS,Planck_bib,bibliography}

\begin{thebibliography}{10}

\bibitem{CMBS4-ScienceBook}
{CMB-S4 Collaboration}.
\newblock {CMB-S4 Science Book, First Edition}, 2016.

\bibitem{henderson/etal:2015}
S.~W.~Henderson et~al.
\newblock {Advanced ACTPol Cryogenic Detector Arrays and Readout}.
\newblock arXiv:1510.02809, October 2015.

\bibitem{suzuki16}
A.~Suzuki et~al.
\newblock {The Polarbear-2 and the Simons Array Experiments}.
\newblock {\em Journal of Low Temperature Physics}, 184:805--810, August 2016.

\bibitem{essinger-hileman/etal:2014}
T.~Essinger-Hileman et~al.
\newblock {CLASS: the cosmology large angular scale surveyor}.
\newblock In {\em Millimeter, Submillimeter, and Far-Infrared Detectors and
  Instrumentation for Astronomy VII}, volume 9153 of {\em Proc. SPIE}, page
  91531I, July 2014.

\bibitem{benson/etal:2014}
B.~A.~Benson et~al.
\newblock {SPT-3G: a next-generation cosmic microwave background polarization
  experiment on the South Pole telescope}.
\newblock In {\em Millimeter, Submillimeter, and Far-Infrared Detectors and
  Instrumentation for Astronomy VII}, volume 9153 of {\em Proc. SPIE}, page
  91531P, July 2014.

\bibitem{grayson16}
J.~A.~Grayson et~al.
\newblock {BICEP3 performance overview and planned Keck Array upgrade}.
\newblock In {\em Millimeter, Submillimeter, and Far-Infrared Detectors and
  Instrumentation for Astronomy VIII}, volume 9914 of {\em Proc. SPIE}, page
  99140S, July 2016.

\bibitem{fraisse13}
A.~A.~Fraisse et~al.
\newblock {SPIDER: probing the early Universe with a suborbital polarimeter}.
\newblock {\em \jcap}, 4:47, April 2013.

\bibitem{gandilo16}
N.~N.~Gandilo et~al.
\newblock {The Primordial Inflation Polarization Explorer (PIPER)}.
\newblock In {\em Millimeter, Submillimeter, and Far-Infrared Detectors and
  Instrumentation for Astronomy VIII}, volume 9914 of {\em Proc. SPIE}, page
  99141J, July 2016.

\bibitem{ebex17a}
The~EBEX Collaboration.
\newblock {The EBEX Balloon Borne Experiment - Optics, Receiver, and
  Polarimetry}.
\newblock {\em ArXiv e-prints}, March 2017.

\bibitem{shariff15}
J.~A. {Shariff}.
\newblock {Polarimetry from the stratosphere with SPIDER and BLASTPol}.
\newblock {\em PhD. Thesis, University of Toronto}, 2015.

\bibitem{planck2016-XLVI}
{Planck Collaboration XLVI}.
\newblock {Planck intermediate results. XLVI. Reduction of large-scale
  systematic effects in HFI polarization maps and estimation of the
  reionization optical depth}.
\newblock {\em Astron. Astrophys.}, 596:A107, 2016.

\bibitem{hou/etal:2013}
Z.~Hou et~al.
\newblock {How massless neutrinos affect the cosmic microwave background
  damping tail}.
\newblock {\em \prd}, 87(8):083008, April 2013.

\bibitem{manzotti/etal:2015}
A.~{Manzotti}, S.~{Dodelson}, and Y.~{Park}.
\newblock {External priors for the next generation of CMB experiments}.
\newblock {\em arxiv/1512.02654}, December 2015.

\bibitem{planck2016-XLVII}
{Planck Collaboration XLVII}.
\newblock {Planck intermediate results. XLVII. Planck constraints on
  reionization history}.
\newblock {\em Astron. Astrophys.}, 596:A108, 2016.

\bibitem{baumann09}
D.~Baumann et~al.
\newblock {Probing Inflation with CMB Polarization}.
\newblock In S.~{Dodelson}, D.~{Baumann}, A.~{Cooray}, J.~{Dunkley},
  A.~{Fraisse}, M.~G. {Jackson}, A.~{Kogut}, L.~{Krauss}, M.~{Zaldarriaga}, and
  K.~{Smith}, editors, {\em American Institute of Physics Conference Series},
  volume 1141 of {\em American Institute of Physics Conference Series}, pages
  10--120, June 2009.

\bibitem{gandilo14}
N.~N.~Gandilo et~al.
\newblock {Attitude determination for balloon-borne experiments}.
\newblock In {\em Ground-based and Airborne Telescopes V}, volume 9145 of {\em
  Proc. SPIE}, page 91452U, July 2014.

\bibitem{datta/etal:2014}
R.~Datta et~al.
\newblock {Horn Coupled Multichroic Polarimeters for the Atacama Cosmology
  Telescope Polarization Experiment}.
\newblock {\em Journal of Low Temperature Physics}, 176:670--676, September
  2014.

\bibitem{datta/etal:2015}
R.~Datta et~al.
\newblock {Design and Deployment of a Multichroic Polarimeter Array on the
  Atacama Cosmology Telescope}.
\newblock {\em ArXiv e-prints}, October 2015.

\bibitem{niemack/etal:2012}
M.~D.~Niemack et~al.
\newblock {Optimizing Feedhorn-Coupled TES Polarimeters for Balloon and
  Space-Based CMB Observations}.
\newblock {\em Journal of Low Temperature Physics}, 167:917--922, June 2012.

\bibitem{henning/etal:2012}
J.~W.~Henning et~al.
\newblock {Feedhorn-coupled TES polarimeter camera modules at 150 GHz for CMB
  polarization measurements with SPTpol}.
\newblock In {\em Millimeter, Submillimeter, and Far-Infrared Detectors and
  Instrumentation for Astronomy VI}, volume 8452 of {\em Proc. SPIE}, page
  84523A, September 2012.

\bibitem{bryan16}
S.~A.~Bryan et~al.
\newblock {A cryogenic rotation stage with a large clear aperture for the
  half-wave plates in the Spider instrument}.
\newblock {\em Review of Scientific Instruments}, 87(1):014501, January 2016.

\bibitem{bicep_III}
{BICEP2 Collaboration}.
\newblock Bicep2 iii: Instrumental systematics.
\newblock {\em The Astrophysical Journal}, 814(2):110, 2015.

\bibitem{louis16}
T.~Louis et~al.
\newblock {The Atacama Cosmology Telescope: Two-Season ACTPol Spectra and
  Parameters}.
\newblock {\em arXiv/1610.02360}, 2016.

\bibitem{austermann_spt_2012}
J.~E.~Austermann et~al.
\newblock Sptpol: an instrument for cmb polarization measurements with the
  south pole telescope.
\newblock volume 8452 of {\em Proc. SPIE}, pages 84521E--84521E--18, 2012.

\end{thebibliography}

\end{document}